\documentclass{iaus}
\usepackage{graphicx}

\usepackage{natbib}
\bibpunct[;~]{(}{)}{;}{a}{}{,}

\newcommand{\aand}{$\alpha$~And}
\newcommand{\teff}{$T_{\rm eff}$}
\newcommand{\cvn}{$\alpha^2$~CVn}

\title[The spots on Ap stars]{The spots on Ap stars}

\author[O. Kochukhov]{Oleg Kochukhov}

\affiliation{Department Physics and Astronomy, Uppsala University, \\ Box
516, SE-75120 Uppsala, Sweden \\email: {\tt oleg.kochukhov@fysast.uu.se}}

\pubyear{2010}
\volume{273}
\pagerange{1--6}
\jname{Physics of Sun and Star Spots}
\editors{A.C. Editor, B.D. Editor \& C.E. Editor, eds.}
\begin{document}

\maketitle

\begin{abstract}
The upper main sequence magnetic chemically peculiar (Ap) stars exhibit a
non-uniform distribution of chemical elements across their surfaces and with
height in their atmospheres. These inhomogeneities, responsible for the
conspicuous photometric and spectroscopic variation of Ap stars, are believed
to be produced by atomic diffusion operating in
the stellar atmospheres stabilized by multi-kG magnetic fields. Here I
present an overview of the current state-of-the-art in understanding 
Ap-star spots and their relation to magnetic fields. In particular,
I highlight recent 3-D chemical spot structure studies and summarize
magnetic field mapping results based on the inversion of the full Stokes vector
spectropolarimetric observations. I also discuss a puzzling new type of spotted stars, HgMn stars, 
in which the formation and evolution of heavy element spots is driven by a
poorly understood mechanism, apparently unrelated to magnetic fields.
\keywords{Polarization, stars: chemically peculiar, stars: magnetic fields, stars: atmospheres}
\end{abstract}

\firstsection 

\section{Introduction}

The early-type magnetic chemically peculiar stars, often referred to as Ap/Bp stars, are main sequence objects found over the entire spectral type range from early B to early F, but most commonly around A0. These stars are not constrained to a particular evolutionary stage \citep{kochukhov:2006} and are found both near the ZAMS in young open clusters \citep[e.g.,][]{folsom:2008} and among the stars at the end of the hydrogen core burning phase \citep{kochukhov:2006b}.

The most conspicuous and unusual property of Ap stars is the presence of strong, globally organized magnetic fields on their surfaces. The field intensity ranges from a well-defined lower threshold of $\approx$\,300~G \citep{auriere:2007}, probably representing the weakest global field able to withstand shearing by the differential rotation, to slightly over 30~kG for a few extreme objects \citep{babcock:1960,elkin:2010}.

The slow rotation of Ap stars, which probably results from an enhanced angular momentum loss during the PMS evolutionary stage \citep{stepien:2000}, the absence of surface convection zone, and the lack of intense mass loss provides ideal conditions for a build up of chemical anomalies by atomic diffusion \citep{michaud:1970}. This slow separation of chemical elements under the influence of radiative pressure and gravity is believed to alter the surface chemistry in many types of stars with extended radiative zones. But it is manifested most strongly in Ap stars, which often exhibit enormous overabundances of rare-earth and other heavy elements as well as a noticeable enhancement of Cr and Fe. The richness of the sharp-lined Ap-star spectra makes these objects the prime astrophysical laboratories for studies of exotic chemical species and for astrophysical verification of the atomic line data \citep{ryabchikova:2006,quinet:2007}.

The presence of a super-equipartition magnetic field, which usually has a topology close to an oblique dipole, breaks the spherical symmetry of the diffusive segregation of chemical elements, leading to the formation of spots and rings of enhanced element abundance \citep{michaud:1981}. The surface magnetic and chemical structure modifies the local energy balance, resulting in a strong UV to optical flux redistribution for the regions of Ap star surface characterized by an overabundance of heavy elements. In combination with the stellar rotation, this yields a periodic variation of the average magnetic field characteristics, line profiles, spectral energy distribution, and brightness in different photometric bands. Both the field structure and the geometry of chemical spots remain stable on the time scale of many decades. This stability allows an accurate analysis of the surface properties of Ap stars through multi-epoch observations and a very precise measurements of rotation rates, even revealing rotational braking for some of these stars \citep{adelman:2001a,mikulasek:2008}.

Rich phenomenological picture of Ap stars has been traditionally addressed with simple empirical models, based on the correlations between photometric variations and the phase curves of the equivalent widths and radial velocities of spectral lines \citep[e.g.][]{polosukhina:2000,leone:2001}, while the magnetic field properties were inferred by fitting low-order multipolar models to measurement of the mean longitudinal magnetic field and the mean field modulus \citep{landstreet:2000}.

Here I review more recent advanced studies of the magnetic field and spots on Ap stars, which have the ambition of developing much more realistic and detailed 3-D models of the chemical element distributions, atmospheric structure, and magnetic field topology. The theoretical basis of these emerging modeling approaches is provided by the developments in the fields of model atmospheres, polarized radiative transfer, and numerical methods of the inverse problem solution, while the observational material -- intensity and polarization spectra at high spectral and time resolution, and often with a coverage of the entire optical domain -- is supplied by the new generation spectrometers and spectropolarimeters at the 4--8-m class telescopes.

\section{Doppler imaging of chemical spots and magnetic fields}

The Doppler imaging (DI) of chemical inhomogeneities on Ap stars was the first application of this powerful remote sensing method to spotted stars \citep{khokhlova:1986}. The basic principle of DI is to use a resolution of the stellar surface provided by the Doppler broadening of line profiles together with the modulation due to stellar rotation for the reconstruction of two-dimensional maps of the stellar surface. For a couple of decades since its invention DI was applied only to a small number of elements \citep[e.g.,][]{hatzes:1991}, ignoring effects of magnetic field on line profiles. At the same, the resulting abundance maps were compared with primitive multipolar models of magnetic field inferred from the longitudinal magnetic field measurements.

The development of the magnetic DI technique by \citet{piskunov:2002a} made it possible to obtain self-consistent maps of abundance distributions and magnetic field using high-resolution polarization spectra \citep{kochukhov:2002b,luftinger:2010}. Moreover, inclusion of the linear polarization spectra in the inversion enabled reconstruction of the vector magnetic field uniquely and without adopting an \textit{a priori} multipolar field parameterization. Ap stars is the only class of spotted magnetically active stars permitting such a sophisticated full Stokes line profile analysis.

\begin{figure}[t]
\begin{center}
\includegraphics[width=\textwidth]{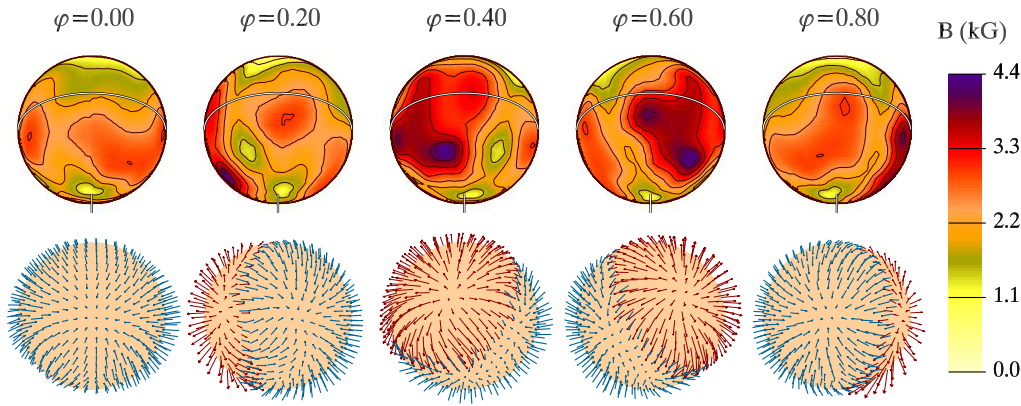}
\caption{Representative results of the modern magnetic Doppler imaging analysis of the
four Stokes parameter spectropolarimetric observations of
a magnetic Ap star. The spherical plots show distribution of the magnetic field strength (\textit{top row}) and 
field orientation (\textit{bottom row}) over the surface of \cvn. Adapted from \citet{kochukhov:2010}.}
\label{fig1}
\end{center}
\end{figure}

As demonstrated by \citet{kochukhov:2004d} for 53~Cam and recently by \citet{kochukhov:2010} for the prototype magnetic Ap star \cvn\ (Fig.~\ref{fig1}), the inclusion of the Stokes $Q$ and $U$ spectra can radically change our view on the field structure of Ap stars. For both stars the overall field topology, determined by the radial field component, is roughly dipolar, but the field intensity distribution is much more complex due to small-scale patches of mainly horizontal magnetic field. These structures cannot be described by any low-order multipolar field and are not detectable with the circular polarization spectra.

Meanwhile non-magnetic abundance mapping of Ap stars has reached maturity in its application to weak-field stars. Efficient numerical algorithms and the use of parallel supercomputers now allow reconstructing distributions for more than a dozen chemical elements \citep{kochukhov:2004e}. These comprehensive horizontal chemical structure studies generally reveal a great diversity of spot geometries, even for the chemical elements with similar spectra and position in the periodic table. A ``typical'' Ap-star chemical spot structure, with the iron peak elements concentrated at the magnetic equator and the rare-earth elements (REE) located at the poles of dipolar field is not confirmed. Only a few chemical elements (Li, O, Eu) show a well-defined spot or ring-like map correlating with the dipolar field component. Most other elements exhibit no such correspondence. It is possible that these elements are preferentially sensitive to the horizontal magnetic field component, which has a very complex topology according to the four Stokes parameter magnetic DI studies.

\section{Vertical stratification of chemical abundances}

Availability of very high quality spectroscopic observations of Ap stars and development of realistic magnetic spectrum synthesis codes \citep{wade:2001} to interpret these observations opened an entirely new dimension of the Ap-spot research. It was recognized that previously dismissed spectral anomalies, such as ionization imbalance, large scatter of abundances derived from lines of different strength and excitation, etc., visible in the Ap-star spectra point to an inhomogeneous vertical distribution of chemical elements \citep{bagnulo:2001,ryabchikova:2002}. A separation of elements into chemically distinct layers was predicted by the theoretical diffusion studies \citep{babel:1992} but was not thoroughly studied observationally until now.

\begin{figure}[t]
\begin{center}
\includegraphics[width=8cm]{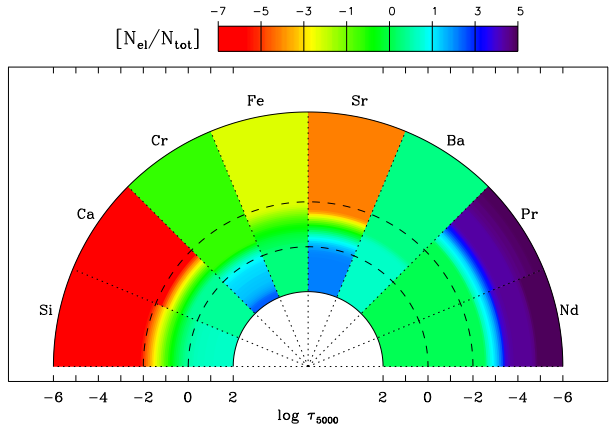}
\caption{Chemical stratification in the atmosphere of the cool Ap star HR\,1217
\citep{shulyak:2009}. Vertical abundance distributions are presented in a
radial plot as a function of the continuum optical depth at $\lambda=5000$~\AA.
Abundances relative to the Sun are given on a logarithmic scale.}
\label{fig2}
\end{center}
\end{figure}

Many recent studies \citep[e.g.,][]{ryabchikova:2006,kochukhov:2006b,kochukhov:2009a,shulyak:2009} showed that all spectral indicators of a given element can be brought into a reasonable agreement with each other by introducing a sharp concentration gradient in the line forming region. Typically, the light and iron peak elements are found to have solar or subsolar abundance in the upper atmospheric layers, above $\log\tau_{5000}$\,=\,0 to $-2$, and a 2--3 dex overabundance at the bottom of the atmosphere. The REEs show an opposite chemical stratification profile, with an overabundance cloud located above $\log\tau_{5000}$\,=\,$-3$. Fig.~\ref{fig2} shows a typical chemical stratification in a cool Ap star, derived in this case for HR\,1217 \citep{shulyak:2009}.

An independent confirmation of the reality of these enormous vertical abundance gradients came from the time-resolved spectroscopic studies of the pulsations in rapidly oscillating Ap (roAp) stars \citep{kochukhov:2006a,ryabchikova:2007b}. All roAp stars show a large discrepancy of the pulsational characteristics between, on the one hand, the light and iron peak element spectral lines, which often show no detectable variability, and REE lines on the other hand, which pulsate with the radial velocity amplitudes of up to several km\,s$^{-1}$. Initially puzzling, this pulsational behaviour is now understood to be a natural consequence of the propagation of magneto-acoustic waves in a chemically stratified atmosphere \citep{khomenko:2009}. The wave amplitude increases rapidly with height, hence pulsations attain the largest amplitude in the uppermost atmospheric layers where REE lines are formed.

The presence of large vertical abundance gradients in many magnetic chemically peculiar stars calls for a revision of the very concept of Ap-star chemical ``spot''. The horizontal abundance maps derived with tradiational DI methods might represent an effective and simplified description of the vertical stratification variation across the surface due to its dependence on the local magnetic field, as anticipated by theoretical models \citep{alecian:2010}. A study aiming to test this hypothesis with a simultaneous horizontal and vertical abundance mapping of chemical elements is currently underway.

\section{Chemical weather in HgMn stars}

\begin{figure}[t]
\begin{center}
\includegraphics[angle=90,width=\textwidth]{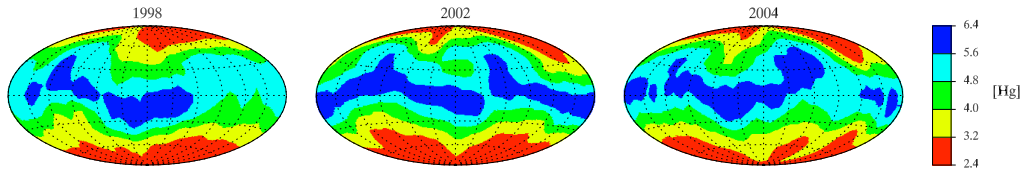}
\caption{Evolution of mercury clouds on the surface of \aand\ \citep{kochukhov:2007b}. 
The DI mercury abundance maps in the Hammer-Aitoff projection show configuration of Hg spots for years 1998, 2002, and 2004. Abundances relative to the Sun are given on a logarithmic scale.}
\label{fig3}
\end{center}
\end{figure}

According to the widely used classification scheme of the upper main sequence chemically peculiar stars \citep{preston:1974} these objects are separated in two distinct groups, depending on the presence of detectable magnetic field on the stellar surface. The objects in the first group, magnetic Ap/Bp stars, show large deviations from solar chemical composition, exhibit strong fossil fields, stable chemical spots, and show rotational variability of the photometric parameters and line profiles. The second group, non-magnetic CP stars, is represented by Am (\teff\,$\le$\,10000 K) and HgMn (\teff\,$\ge$\,10000 K) stars, which are characterized by moderate chemical anomalies and were believed to show neither photometric nor spectroscopic variability. Spectropolarimetric surveys of Am and HgMn stars \citep{shorlin:2002} and weak-field Ap stars \citep{auriere:2007} reinforced the notion of this magnetic dichotomy by showing that the minimum magnetic field found in Ap stars significantly exceeds the upper magnetic field strength limits established for bright Am and HgMn stars. In this context, a strong magnetic field was thought to be a necessary prerequisite for the formation of chemical abundance spots on early-type stars.

This understanding of the relation between magnetism and spot formation on early-type stars has proved to be wrong. A series of recent spectroscopic studies demonstrated the presence of surface chemical inhomogeneities in several HgMn stars \citep[][Korhonen et al., this meeting]{adelman:2002,kochukhov:2005b,folsom:2010,briquet:2010}. These non-uniform abundance distributions typically have a lower contrast than those found on the magnetic Ap stars of similar \teff\ and are observed only for a small number of chemical species, such as Ti, Y, Sr, Pt, and Hg, showing significant anomaly of the average abundance compared to the solar chemical composition.

The discovery of spots on HgMn stars revived the discussion of their magnetic status. Although no conclusive magnetic field detections were reported for these stars by studies using previous generation spectropolarimeters \citep{shorlin:2002}, the precision of these magnetic field measurements was not high enough to exclude the presence of $\sim$\,100~G fields, which can still play an important role in the spot formation process. More recent sensitive searches for the magnetic field signatures using new generation spectropolarimeters targeted known HgMn stars with spots, \aand\ \citep{wade:2006} and AR~Aur \citep{folsom:2010}, finding no longitudinal field stronger than $\approx$\,10~G for the former star and $\approx$\,30~G for the latter. Finally, Makaganiuk et al. (submitted) carried out a comprehensive magnetic survey of over 40 HgMn stars with the new HARPSpol instrument at the ESO 3.6-m telescope. None of their targets, which included several spotted HgMn stars, showed any evidence of magnetic field. For the majority of stars this survey achieved an upper limit of 10--20~G for the longitudinal magnetic field, and as low as 1--3~G for several sharp-lined stars. Analysis of the circular polarization profiles at the resolving power of $>$\,$10^5$ also do not reveal complex magnetic fields similar to those seen in active late-type stars. These results confirm the non-magnetic status of HgMn stars and suggest that the spot formation on their surfaces is unrelated to magnetic field.

A new light on the puzzling nature of chemical inhomogeneities in HgMn stars was shed by the discovery of temporal evolution of the spot topology in \aand\ by \citet{kochukhov:2007b}. In this study the Hg spots were followed over a time span of 7 years with the spectra of exquisite quality ($S/N$\,$\sim$\,1000), revealing slow changes of the surface structure both in the original data and in the resulting Doppler images (Fig.~\ref{fig3}). Later, a faster evolution of chemical spots was claimed for the HgMn star HD\,11753 by \citet{briquet:2010} based on Doppler maps obtained for two epochs separated by only 65 days.

It is clear that HgMn stars represent an interesting new type of spotted stars, in which magnetic field is not instigating the chemical spot formation. In this respect the spots on HgMn stars are fundamentally different from both the stable abundance inhomogeneities found on the early-type stars with fossil magnetic fields and the temperature spots associated with complex field topologies on late-type active stars. Instead, the spots on HgMn stars probably form and evolve under the influence of time-dependent atomic diffusion processes and hydrodynamical instabilities in the upper layers of chemically stratified atmosphere \citep{kochukhov:2007b}. Detailed theoretical diffusion models are needed to understand the origin of this newly discovered, remarkable spot formation phenomenon.

\end{document}